\documentclass[12pt]{article}

\usepackage[paper=a4paper, margin=2cm]{geometry}
\usepackage[russian,english]{babel}
\usepackage[cp1251]{inputenc}
\usepackage{amsmath}

\usepackage[logos]{amsbib}

\begin{document}
%%%%%%%%%%%%%%%%%%%%%%%%%%%%%%%%%%%%%%%%%%%%%%%%%%%%%%%%%%%%%%%%%%%%%%%%%%%%%%%%%%%%

\date{}
\renewcommand{\refname}{References}

\author{A.~Yu.~Samarin%, samarinay@yahoo.com
}
\title{Motion of macroscopic bodies in terms of quantum evolution}

\maketitle

\centerline{Samara Technical State University, 443100 Samara, Russia}
\centerline{}

\abstract{ The mechanical motion of the center of mass of a system in physical space is described by using Feynman's representation of quantum evolution. It is shown that alone unitary evolution of a closed system of quantum particles without decoherence and collapse phenomena generates the classical mechanical motion of the system's center of mass, when the number of these particles go to infinity. The approach, introduced here, allows considering a mechanical system with the classical properties, as the set of particles obeying the quantum laws exclusively. The rigorous mathematical image of macroscopic objects in term of quantum mechanics allows considering any non-relativistic processes involving both quantum and macroscopic objects; this is necessary, first of all, to describe quantum measurements.

{

%%%%%%%%%%%%%%%%%%%%%%%%%%%%%%%%%%%%%%%%%%%%%%%%%%%%%%%%%%%%%%%%%%%%%%%%%%%%%%%%%%%%

\begin {center} \textbf{1. Introdution}\\
\end {center}

Since Schr{\"{o}}dinger formulated the problem of the description of the motion of a macroscopic body for the first time in 1935 \cite{bib:1}, it has become one of the most widely discussed problems in quantum mechanics. Many attempts, based on different interpretations of quantum mechanics, have been made to overcome the difficulties connected with this problem \cite[pp.xvii--xxxix]{bib:2},  \cite{bib:3, bib:4}. However, there is no mathematical representation of the mechanical motion of the macroscopic object in terms of quantum mechanics~\cite{bib:5}. Not only does this fact limits the generality of the quantum theory, but it does not also allow to describe the reduction of the wave function~\cite[p.351]{bib:6},~\cite[p.36]{bib:7}, caused by quantum measurements, in the form of a causal chain of events~\cite{bib:8}. Ability to present the macroscopic body as a collection of a large number of the quantum objects is necessary to start the measuring process study having at least some chance of success~\cite{bib:9}. The collapse conception as a unique nonlocal physical phenomenon ~\cite{bib:10,bib:11,bib:12,bib:13,bib:14} is necessary in order to find out the domain of applicability of the notion "observable" and, perhaps, to point out the mathematical objects that are suitable for using outside this domain ~\cite[pp.163--195]{bib:15} (these quantities, of course, must be expressed in terms of observables under certain conditions). Such an approach would likely allow to study the processes, detail description of which is impossible using observables due to uncertainty principle~\cite{bib:16,bib:17}. In addition, it would likely allow to describe the evolution of the quantum system in the form of mechanical motion in physical space~\cite{bib:18} and introduce the spacetime notion for quantum objects, what is necessary for the creation of quantum gravitation~\cite{bib:19,bib:20}.

The macroscopic path of the particle corresponding to the least action principle has been obtained as a classical limit of the path integral~\cite[pp.29,30]{bib:15}. However this does not solve the problem of macroscopic body description: first of all, the mechanism of localization of the wave function has not been shown and, secondly, the macroscopic body cannot be considered as a particle if we want to describe the quantum system in general case.

In order to introduce the mathematical image of a macroscopic body, laws of quantum mechanics, first of all, we have to determine the macroscopic body in a suitable manner. In general, the macroscopic body is a system of a vast number of particles. Individual motion any of these particles obeys the quantum laws. Therefore, a quantum system can posses macroscopic properties only as a whole. Moreover, in accordance with~\cite[p.4]{bib:7} the quantum equations will be in close correspondence with the equation of classical mechanics, in other words, dynamics laws of classical mechanical motion have to be a direct consequence of quantum laws. Taking this fact into account, the following statement can be made: the macroscopic body is a system of quantum particles that the center of mass moves in accordance with the laws of classical mechanics. Namely the center of mass as a formal mathematical notion characterizing a quantum system in the form of a unified object can possess classical properties.

 Mechanical motion of the center of mass in physical space is the subject of consideration in this paper

\begin {center} \textbf{\textbf{2.The roots of the problem}}\\
\end {center}

The first serious problem of description of macroscopic objects by means of conventional quantum mechanics is following: the center of mass of a macroscopic body is always localized in space, whereas any quantum system in accordance with this theory is localized only immediately after the collapse. Quantum evolution within the framework of this theory does not result in localization of the center of mass of the system consisting of any large numbers of particles. In order to overcome this difficulty in~\cite{bib:4,bib:21,bib:22} an accessory postulate has been added to the conventional quantum mechanics postulates. This postulate asserts, that there is a spontaneous localization of quantum particles taking place in the macroscopic body practically all the time. Nevertheless, even so meaningful postulate does not allow to deduce the classical law of the center of mass mechanical motion starting from the quantum mechanics dynamics.

Supposing that the collapse is a consequence of the measuring process and taking into account that the physics of classical systems can be treated without ever referring to measurement~\cite{bib:23}, the mechanical motion of the center of mass will be considered solely basing on the unitary evolution law, taken in a suitable form.

The least action principle determinimg the mechanical motion of the center of mass of a macroscopic body is formulated for physical space. To deduce this principle from quantum evolution, the latter has to be described correspondingly, i.e. in the spatio-temporal form. A major step towards the spatio-temporal description of unitary quantum evolution was taken by Dirac and Feynman~\cite{bib:24,bib:25}.

 \begin {center} \textbf{\textbf{2. The Law of Dynamics}}\\
\end {center}

As we consider the system of quantum particles then, first of all, let us determine the notion of a quantum particle in the spatio-temporal form. The quantum particle is an indivisible quantum object, that is transformed into a mass point as the result of the collapse. Formally this means, that coordinates wave function of the quantum particle becomes $\delta$-function as the result of the space localization in physical space. Then the particles can be considered as matter fields or collections of matter fields. This depends on whether the corresponding wave functions are nondegenerate or degenerate. In any case, it is sufficient to consider the nondegenerate case (This does not lead to the loss of generality of the consideration due to superposition principle).

 The mechanical motion of these matter fields (continua) can not be described in terms of the Hilbert space: the set of basis vectors of the Hilbert space is countable, whereas the coordinates set of physical space has cardinality of continuum. This means that the spatial variables are not observables and the coordinate form of the wave equation have to be considered as the dynamic law of mechanical motion in physical space. As we consider the path of the center of mass, the wave equation in the integral form is more suitable than the differential wave equation.

Consider the evolution of the system which consist of $n $ quantum particles in the form of one-dimensional mechanical motion of these particles. Let $x^{1}...x^{n}$ be the space coordinates of these particles. Take as a postulate, the dynamic law of this motion in the form of the integral wave equation ~\cite{bib:25},~\cite[pp.163--195]{bib:15}:
\begin{eqnarray}\label{eq:math:ex1} 
\Psi_{t_{2}}(x_{2}^{1},  \dots  ,x_{2}^{n})
    &=&\int...\int K_{t_{2},t_{1}}(x_{2}^{1}, \dots ,x_{2}^{n},x_{1}^{1}, \dots ,x_{1}^{n})\nonumber \\
 & &\times \Psi_{t_{1}}(x_{1}^{1}, \dots ,x_{1}^{n})\,dx^1_1\cdots\,dx^n_1,
\end{eqnarray}
where $\Psi_{t_{2}}(x_{2}^{1}, \dots ,x_{2}^{n}) $, $\Psi_{t_{1}}(x_{1}^{1}, \dots ,x_{1}^{n}) $ are the wave functions of the system at the time $t_{2} $ and the time $t_{1}<t_{2} $ correspondingly; $K_{t_{2},t_{1}}(x_{2}^{1}, \dots ,x_{2}^{n},x_{1}^{1}, \dots ,x_{1}^{n}) $  is the kernel of the integral evolution operator. The superscript of the spatial variable denotes the number of the particle, the subscript denotes the instant of time. In accordance with the assumption that the quantum particles are matter fields, the spatial variables can be interpreted as the space coordinates of the individual particles of these continua. The wave function and the kernel of the integral evolution operator depend on time parametrically. Further, instead of the term "kernel", the term "transition amplitude" will be used (this amplitude formally corresponds with the transition between the states having spatial localization).

Let the configuration space has the coordinate axes corresponding with the particles of the system. Then, the transition amplitude has the form of continual integral ~\cite{bib:26} in this configuration space. By $\Gamma $ denote an arbitrary virtual path in this space. Thus, we have
\begin{eqnarray*}
K_{t_{2},t_{1}}(x_{2}^{1}, \dots ,x_{2}^{n},x_{1}^{1}, \dots ,x_{1}^{n})=\int \exp{\frac{i}{\hbar}S_{1,2}[\Gamma]}\,[d\Gamma].
\end{eqnarray*}
 The subscript of the action functional denotes the positions of the system in space and time. Let $\gamma $ be a virtual path of a quantum particle in space. Substituting $\gamma $ for $\Gamma $ in the last expression, we get
\begin{eqnarray}\label{eq:math:ex2}
K_{t_{2},t_{1}}(x_{2}^{1}, \dots ,x_{2}^{n},x_{1}^{1}, \dots ,x_{1}^{n})&&\nonumber\\=\int...\int \exp{\frac{i}{\hbar}S_{1,2}^{\Sigma}[\gamma^{1}, \dots ,\gamma^{n}]}\,[d\gamma^{1}]\cdots[d\gamma^{n}],
\end{eqnarray}
where
\begin{eqnarray}\label{eq:math:ex3}
S_{1,2}^{\Sigma}[\gamma^{1}, \dots ,\gamma^{n}]=\sum\limits_{j=1}^{n}\int\limits_{t_{1}}^{t_{2}}\Biggl(m^{j}\frac{(v^{j})^{2}}{2}-U^{j}(x^{j})&&\nonumber\\-\sum\limits_{k=1,k\neq j}^{n}U^{jk}(x^{j},x^{k})\Biggl)\,dt
\end{eqnarray}
is the sum of all the action functionals of the particles forming the system. In this expression  $\frac{m^{j}(v^{j})^{2}}{2} $ is the kinetic energy of the particle $j$ for the path $\gamma^{j}$; $U^{j}(x^{j}) $ is the potential energy of this particle in external fields for the same path; $\sum\limits_{k=1,k\neq j}^{n}U^{jk}(x^{j},x^{k}) $ is the interaction energy of the particles $j$ and $k$ of the system for the paths $\gamma^{j}$ and $ \gamma^{k}$ correspondingly.

\begin {center} \textbf{\textbf{3.  Motion of the Center of Mass}}\\
\end {center}

Consider the motion of the center of mass of the system. Each virtual path of the system in the configuration space determines a unique virtual path of the center of mass in the form
\begin{eqnarray*}
X_{\Gamma}(t)=\frac{\sum\limits_{j=1}^{n}m^{j}x_{\Gamma}^{j}(t)}{\sum\limits_{j=1}^{n}m^{j}}.
\end{eqnarray*}
Here, the path $x_{\Gamma}^{j}(t) $ of an individual particle corresponds with the path $\Gamma $. In this case the total functional~(\ref{eq:math:ex3}) is the sum of the functionals corresponding with the motion of the center of mass and the relative motion of the particles of the system:
 \begin{eqnarray*}
 S_{12}^{\Sigma}=S_{12}^{C}+S_{12}^{R},
\end{eqnarray*}
where
\begin{eqnarray*}
 S_{1,2}^{C}=\int\limits_{t_{1}}^{t_{2}}\Biggl(M\frac{(V)^{2}}{2}-\sum\limits_{j=1}^{n}U^{j}(X,\xi^{j})\Biggl)\,dt;
\end{eqnarray*}
\begin{eqnarray*}
 S_{1,2}^{R}=\sum\limits_{j=1}^{n}\int\limits_{t_{1}}^{t_{2}}\Biggl(m^{j}\frac{(\dot{\xi}^{j})^{2}}{2}-\sum\limits_{k=1,k\neq j}^{n}U^{jk}(\xi^{j},\xi^{k})\Biggl)\,dt;
\end{eqnarray*}
$V=\dot{X}$ --- the velocity of the center of mass; $\xi^{j} $ --- the coordinate of the relative movement of the quantum particle $j $; $M $ --- the mass of the system. Using these notation, instead of ~(\ref{eq:math:ex2}) we get:
\begin{eqnarray*}
 K_{t_{2},t_{1}}(X_{2},\xi_{2}^{1}, \dots ,\xi_{2}^{n-1},X_{1},\xi_{1}^{1}, \dots ,\xi_{1}^{n-1})\\
    =\int \,[dX(t)]\int...\int \exp{\frac{i}{\hbar}\biggl( S_{1,2}^{C}+S_{1,2}^{R}}\biggl)\\ \times\,[d\xi^{1}(t)]\cdots[d\xi^{n-1}(t)].
\end{eqnarray*}
For us to be able to represent a mechanical system as a matter point, we have to suppose that the potential energy of the particles of this system  in the external field does not depend on the space coordinates of the relative movement, i.e. $U^{j}(X,\xi^{j})\approx U^{j}(X) $. In addition, the kinetic energy of the center of mass $\frac{MV^{2}}{2} $ does not depend on these coordinates. Then, the motion of the center of mass does not depend on the relative movement. Thus, we have the following equation for the transition amplitude of the center of mass:
\begin{eqnarray*}
 K_{t_{2},t_{1}}(X_{2},X_{1})\\=\int \exp\Biggl({\frac{i}{\hbar}\int\limits_{t_{1}}^{t_{2}}\biggl(T(V(t))-U(X(t))\biggl)dt\Biggl)}\,[dX(t)].
\end{eqnarray*}
In order to review the path of a macroscopic body, it is necessary to estimate the contribution of different virtual paths to the above continual integral.

\begin {center} \textbf{\textbf{4. The Macroscopic body Path}}\\
\end {center}

There is a formal mathematical procedure which transforms the complex path integral into the real form ~\cite{bib:27}. According to this procedure, the time variable is transformed into the complex form $t=\tau\exp i\varphi $. Then the path integral is considered for the imaginary negative time ($\varphi=-\frac{\pi}{2}\Rightarrow t=-i\tau $):
 \begin{eqnarray*}
K_{\tau_{2},\tau_{1}}(X_{2},X_{1})\\=\int \exp\Biggl({-\frac{1}{\hbar}\int\limits_{\tau_{1}}^{\tau_{2}}\biggl(T(V(\tau))+U(X(\tau))\biggl)d\tau\Biggl)}\,[dX(\tau)],
\end{eqnarray*}
 where $T(V(\tau))=\frac{M}{2}\Biggl(\frac{dX(\tau)}{d\tau}\Biggl)^{2} $ is  the kinetic energy expressed as the function of time module. Using the previous expression and equation~(\ref{eq:math:ex1}), we obtain
\begin{eqnarray*}
\Psi_{\tau_{2}}(X_{2})\\=C\int\Biggl(\int\exp\biggl(-\frac{1}{\hbar}S_{12}^{E}[X(\tau)]\biggl)[X(\tau)]\Biggl) \Psi_{_{1}}(X_{1})\,dX_1.
\end{eqnarray*}
Here $S_{12}^{E}[X(\tau)] $ is the Euclidean action functional, $C $ --- the normalization factor. Denote by $S_{12}^{E}(X_{1},X_{2}) $ the action functionals that have initial and final positions of the center of mass assigned. Let $S_{12}^{E,min}(X_{1},X_{2}) $ be the least of these action functional. Then we have
\begin{eqnarray*}
 \Psi_{\tau_{2}}(X_{2})
 =C\int\exp\biggl(-\frac{1}{\hbar}S_{12}^{E,min}(X_{1},X_{2})\biggl)\\\times\Biggl(\int\exp\biggl(-\frac{1}{\hbar}\Delta S_{12}^{E}[X(\tau)]\biggl)[X(\tau)]\Biggl) \Psi_{_{1}}(X_{1})\,dX_1.
\end{eqnarray*}
Here $\Delta S_{12}^{E}[X(\tau)]=S_{12}^{E}[X(\tau)]- S_{12}^{E,min}(X_{1},X_{2})$.
   Let $s_{12}^{E,min} $ be the least of all the action functional $S_{12}^{E,min}(X_{1},X_{2}) $ having different coordinates. Then, we obtain
\begin{eqnarray*}
 \Psi_{\tau_{2}}(X_{2})=C\exp\biggl(-\frac{1}{\hbar}s_{12}^{E,min}\biggl)\\
  \times\int\exp\biggl(-\frac{1}{\hbar}\Delta S_{12}^{E,min}(X_{1},X_{2})\biggl)\\\times\Biggl(\int\exp\biggl(-\frac{1}{\hbar}\Delta S_{12}^{E}[X(\tau)]\biggl)[X(\tau)]\Biggl) \Psi_{_{1}}(X_{1})\,dX_1.
\end{eqnarray*}
  Here $\Delta S_{12}^{E,min}(X_{2},X_{1})=S_{12}^{E,min}(X_{2},X_{1})- s_{12}^{E,min}$. The exponential factor $\exp\biggl(-\frac{1}{\hbar}s_{12}^{E,min}\biggl) $ does not depend on space coordinates and can be included in the normalization factor $C'$. Consequently,
\begin{eqnarray*}
  \Psi_{\tau_{2}}(X_{2})=C^{\prime}\\
 \times\int\exp\biggl(-\frac{1}{\hbar}\Delta S_{12}^{E,min}(X_{1},X_{2})\biggl)\\\times\Biggl(\int\exp\biggl(-\frac{1}{\hbar}\Delta S_{12}^{E}[X(\tau)]\biggl)[X(\tau)]\Biggl) \Psi_{_{1}}(X_{1})\,dX_1.
\end{eqnarray*}
A macroscopic body contains an infinite number of quantum particles. Therefore, for any finitesimal time interval, the center of mass of this system has such Euclidian action that $S_{12}>>\hbar $ on any path. This situation can be formally expressed as $\hbar\rightarrow 0 $. In this case the integral measure of all sets of the virtual paths for that $\Delta S_{12}^{E,min}(X_{1},X_{2})\neq0$ or $\Delta S_{12}^{E,min}[X(\tau)]\neq0$ is equal to zero because of  $\lim\limits_{\hbar\rightarrow 0}\exp\biggl(-\frac{1}{\hbar}\Delta S_{12}^{E,min}(X_{1},X_{2})\biggl)=0 $ and $\lim\limits_{\hbar\rightarrow 0}\exp\biggl(-\frac{1}{\hbar}\Delta S_{12}^{E}[X(\tau)]\biggl)=0 $. Therefore we have a unique path that determines the last path integral. The spacial part of the wave function $\Psi_{\tau_{2}}(X_{2})$ is the delta-function $\delta(X_{2}-X_{2}^{min})$. The path $X^{min}(\tau) $ corresponds with the least Euclidian action  $s_{12}^{E,min} $. If to go back to the real time, using the analytic continuation, we obtain usual principle of least least action and the spacial wave function $\Psi_{t_{2}}(X_{2})=\delta(X_{2}-X_{2}^{min})$.

\begin {center} \textbf{\textbf{4. Conclusion}}\\
\end {center}

Thus, the classical mechanical motion of a system is the direct result solely of the quantum evolution of the system (without collapse). The macroscopic motion of the center of mass does not depend on the wave function of the relative motion of the particles forming the system. Therefore decoherence phenomenon does not affect the motion of the center of mass  and, thus, cannot not define the system as a macroscopic object.

This paper approach allows us to calculate the deviation scope of the quantum particle system behavior from the classical one, namely, it enables us to find the probability of the center of mass deviation from the path corresponding to the least action principle. This would give an opportunity to establish clearly the connection between quantum and classical descriptions. This connection can be verified experimentally using mesoscopic systems.

In order to describe the collapse as a physical phenomenon it is necessary to consider interaction of the quantum object with the measurer, using the integral wave equation. The wave function collapse have to be the result of the initiation of a macroscopic process in the measuring instrument. This process can be expressed mathematically strictly by the center-of-mass motion of the measuring instrument particles.\\

\vfill\eject

\end{document}